\documentclass[epj-spec]{svjour}

\usepackage{graphics}

\begin{document}

\title{Sensitivity to parameters of STIRAP in a Cooper Pair Box}

\author{G. Mangano\inst{1,2}\fnmsep\thanks{\email{giuseppe@femto.dmfci.unict.it}}\and J. Siewert\inst{1,2} \and G. Falci\inst{1}}

\institute{MATIS CNR-INFM \& Dipartimento di Metodologie Fisiche e Chimiche per l'Ingegneria (DMFCI), Universit\'{a} di Catania, I-95125 Catania, Italy  \and Institute f\"{u}r Theoretische Physik, Universtit\"{a}t Regensburg, D-93040 Regensburg, Germany}

\abstract{
The rapid experimental progress in the field of superconducting nanocircuits gives rise to an increasing quest for advanced quantum-control techniques for these macroscopically coherent systems. Here we demonstrate theoretically that stimulated Raman adiabatic passage (STIRAP), a well-established method in quantum optics, should be possible with the quantronium setup of a Cooper-pair box. We find the parameters which optimize the procedure and show how the scheme appears to be robust against decoherence and should be realizable even with the existing technology.
} 

\maketitle

\section{Introduction}
\label{intro}
During the past decade the successful development of nanotechnology has made it possible to observe macroscopic quantum phenomena in superconducting nanocircuits based on Josephson junctions. These circuits behave like ``artificial atoms", and phenomena, traditionally part of nuclear magnetic resonance (NMR), quantum optics and cavity quantum electrodynamics, have been successfully implemented in such systems. Rabi oscillations and Ramsey interferometry  \cite{Nakamura99,Vion02,Chiorescu03}, the detection of geometric phases \cite{Falci00}, two coupled artificial atoms \cite{Yamamoto03,Majer05}, artificial atoms coupled to electromagnetic resonators \cite{Wallraff04,Chiorescu04}, cooling techniques \cite{Martin04}, an analogue of electromagnetically induced transparency \cite{Murali04} and adiabatic passage in superconducting nanocircuits \cite{Amin03,Siewert04,Siewert05,Paspalakis04,Liu05} are examples first realized in atomic systems that have also been recently demonstrated with superconducting quantum circuits.

One of the strongest motivations of these studies is to apply more complex control techniques of the dynamics in a nanodevice. Advanced control has been proposed as a tool to achieve dynamical decoupling from solid-state noise~\cite{bang-bang}.
Compared to quantum optics and NMR, in these systems coupling energies larger than in atomic physics are easily achieved, thus reducing typical time scales for operations. Flexibility in the design offers several solutions for tuning couplings, allowing in principle to implement the Hamiltonian and to achieve any desired state transformation. On the other hand, the macroscopic nature of the system and the effects of solid state noise impose obstacles requiring more than a mere translation from quantum optics to solid state.

A particularly interesting technique in quantum optics is the so-called stimulated Raman adiabatic passage (STIRAP), developed by Bergmann and co-workers \cite{Bergmann98}. This method represents one of the most efficient coherent population transfer schemes between two (or more) quantum states known in quantum optics. It has many applications in such diverse areas as chemical-reaction \cite{Dittmann92}, laser-induced cooling \cite{Kulin97}, single photon generation in atom-cavity system \cite{Parkins93,Hennrich00} and represents a building block of non-abelian geometric quantum computation procedures \cite{Unanian99}.

In this work we will demonstrate the application of the STIRAP technique to a single Cooper-pair box in the charge-phase regime (the so-called quantronium) \cite{Vion02}. Although one can apply this procedure to different regimes and setups of superconducting nanocircuits \cite{Paspalakis04,Liu05,Siewert04,Siewert05}, we have chosen the quantronium because it is very similar to the atom-laser system in quantum optics and it has been extensively studied with respect to its decoherence properties. Actually the main problem in nanodevices is the trade-off between efficient coupling to the driving field and protection against noise. In this respect quantronium offers a uniquely convenient design, where tunability is exploited to obtain selective and relatively strong coupling to the fields allowing to perform STIRAP before decoherence takes place.

In Section 2 we will describe the STIRAP protocol in the quantronium. In Section 3 we will find the parameters which optimize the efficiency of the population transfer and finally in Section 4 we will  discuss the feasibility of the protocol in the presence of decoherence.

\section{STIRAP in the quantronium}
\label{sec:1}
The STIRAP technique in quantum optics is based on a three state linkage forming a $\Lambda$ pattern of two hyperfine ground states $|g\rangle$ and $|u\rangle$ coupled to an excited state $|e\rangle$ by classical laser fields $A_{\mathrm{g}}(t) \cos{\omega_{\mathrm{g}}t}$, 
$A_{\mathrm{u}}(t) \cos{\omega_{\mathrm{u}}t}$ \cite{Bergmann98,Scully}. Applying the unitary transformation diag($1,e^{i\omega_u t},e^{i\omega_g t}$) in a basis $\{|e\rangle,|u\rangle,|g\rangle\}$ and the rotating wave approximation, the three-level Hamiltonian reads ($\hbar=1$)
\begin{equation}
\widetilde{H}\  = \ \delta_{\mathrm{g}}|e\rangle\langle e|+ (\delta_{\mathrm{g}}-\delta_{\mathrm{u}})|u\rangle\langle u|
           + \frac{1}{2}(A_{\mathrm{u}}(t)|e\rangle\langle u|
           +A_{\mathrm{g}}(t)|e\rangle\langle g|+{\rm h.c.})\ \ 
\end{equation}
where $\delta_{\mathrm{g}}=E_{\mathrm{e}}-E_{\mathrm{g}}-\omega_{\mathrm{g}}$ and $\delta_{\mathrm{u}}=E_{\mathrm{e}}-E_{\mathrm{u}}-\omega_{\mathrm{u}}$ are the detunings and $A_{\mathrm{u}}(t)$, $A_{\mathrm{g}}(t)$ the slowly-varying Rabi frequencies. 

An essential condition for STIRAP is that there be two photon-resonance between states $|g\rangle$ and $|u\rangle$ meaning 
\begin{equation}
 \delta \ \equiv \ \delta_{\mathrm{g}}-\delta_{\mathrm{u}} \ = \ 0
\end{equation}
Under this condition, setting $\Delta=\delta_{\mathrm{g}}=\delta_{\mathrm{u}}$, the Hamiltonian becomes
\begin{equation}
\label{stirap_hamilt}
\widetilde{H}\  = \ \Delta|e\rangle\langle e|+
           + \frac{1}{2}(A_{\mathrm{u}}(t)|e\rangle\langle u|
           +A_{\mathrm{g}}(t)|e\rangle\langle g|+{\rm h.c.})\ \ 
\end{equation}
The peculiarity of this Hamiltonian is the presence of a null instantaneous eigenvalue, whose associated adiabatic state is essential for the population transfer mechanism. It has no component of the excited state $|e\rangle$, being a coherent superposition of the states $|g\rangle$ and $|u\rangle$ only and can be written as
\begin{equation}
\label{dark}
|\phi^{0}\rangle = |g\rangle \ \cos{\theta(t)} - |u\rangle \ \sin{\theta(t)}
\end{equation}
where the time-dependent mixing angle is defined as 
\begin{equation}
\tan{\theta(t)}=\frac{A_{\mathrm{g}}(t)}{A_{\mathrm{u}}(t)}
\end{equation}
The latter state cannot decay by spontaneous emission from $|e\rangle$ and is usually called dark state. From Eq. (\ref{dark}) it can be seen that by slowly varying the coupling amplitudes $A_{\mathrm{u}}(t)$, $A_{\mathrm{g}}(t)$ the dark state can be rotated in the two dimensional subspace spanned by $|g\rangle$ and $|u\rangle$. STIRAP is based on tying the state vector to the dark state. Population transfer from the state $|g\rangle$ to the state $|u\rangle$ can be achieved by applying the so-called counterintuitive scheme \cite{Bergmann98}: the system is prepared in the state $|g\rangle$ at vanishing couplings and the state vector coincides with the dark state; then the field $A_{\mathrm{u}}$ is turned on while $A_{\mathrm{g}}$ is kept zero. Finally, by slowly switching $A_{\mathrm{u}}$ off while $A_{\mathrm{g}}$ is switched on, the dark state evolves from the bare state $|g\rangle$ to the target state $|u\rangle$. If the adiabatic condition is fulfilled, the state vector will ``follow" the dark state and the population transfer is achieved. 

\begin{figure}
\centering
\resizebox{0.4\textwidth}{!}{\includegraphics{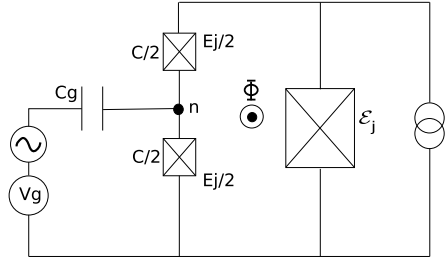}}
\caption{Idealized circuit diagram of the quantronium. This quantum-coherent circuit consists of a Cooper pair box island (black node) delimited by two small Josephson junctions (crossed boxes) in a superconducting loop. The bigger junction with Josephson energy $\varepsilon_{\mathrm{J}}$ is used for readout. The offset charge of the island can be tuned with the gate voltage $V_{\mathrm{g}}$ ($C_{\mathrm{g}}\ll C $), while the Josephson energy of the junctions can be controlled via the magnetic flux $\Phi$. Here we choose $\Phi=0$ and $E_{\mathrm{J}}/E_{\mathrm{C}}=1.318$.}
\label{quantronium}
\end{figure}

\begin{figure}[ht]
\centering
\resizebox{0.5\textwidth}{!}{\includegraphics{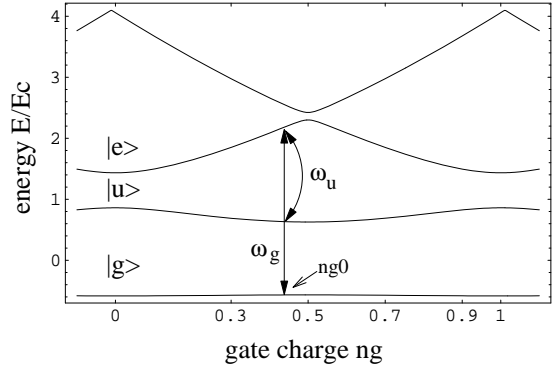}}
\caption{The four lowest energy levels of the quantronium as a function of the gate charge $n_g$. At the chosen working point $n_{g0}$, STIRAP can be performed between the three lowest energy levels $|g\rangle$, $|e\rangle$ and $|u\rangle$ with resonance frequencies $\hbar\omega_g=E_e-E_g$ and $\hbar\omega_u=E_e-E_u$.}
\label{spectrum}
\end{figure}

Our aim now is to implement the Hamiltonian (\ref{stirap_hamilt}) in the quantronium and realize adiabatic population transfer. The quantronium circuit \cite{Vion02} is showed in Fig. \ref{quantronium}. It consists in a superconducting loop interrupted by two adjacent tunnel junctions with Josephson energies $E_{\mathrm{J}}/2$ and by the larger readout junction (with $\varepsilon_{\mathrm{J}} \gg E_{\mathrm{J}}$).
The two small junctions define the superconducting island of the box, whose total capacitance is $C$ and  charging energy $E_{\mathrm{C}}=(2e)^2/ 2 C$. Except at readout, the circuit can be modelled in the laboratory frame by the following Hamiltonian

\begin{equation}
\label{quantronium_hamilt}
H = \sum_n E_{\mathrm{C}} [n-n_g(t)]^2|n\rangle\langle n| 
- \frac{E_{\mathrm{J}}}{2} (|n\rangle\langle n+1| +\mathrm{h.c.})
\end{equation}
where $\{|n\rangle\}$ are eigenstates of the number operator $\hat{N}$ of extra Cooper pairs on the island and $n_{\mathrm{g}}(t)= C_{\mathrm{g}} V_{\mathrm{g}}/(2e)$ is the reduced gate charge. The Cooper pair number $\hat{N}$ and the superconducting phase difference across the readout junction are the degrees of freedom of the circuit. A dc voltage $V_{\mathrm{g}}$ applied to the gate capacitance $C_{\mathrm{g}}$ and a dc current applied to a coil  producing a flux  in the circuit loop, tune the quantum energy levels. Here we choose $\Phi=0$ and a ratio $E_{\mathrm{J}}/E_{\mathrm{C}}=1.3$ ($E_{\mathrm{C}}=0.66 K_{\mathrm{B}} K$, $E_{\mathrm{J}}=0.87 K_{\mathrm{B}} K$ as reported in Ref. \cite{Vion02}). Being $E_{\mathrm{J}}\simeq E_{\mathrm{C}}$, neither $\hat{N}$ nor the phase different is a good quantum number. The box thus has discrete quantum states which are quantum superpositions of several charge states with different $n$. Diagonalizing the Hamiltonian (\ref{quantronium_hamilt}) for a certain dc bias voltage $n_{\mathrm{g}}(t)=n_{\mathrm{g0}}$, one finds

\begin{equation}
H = \sum_i E_{i} |\phi_i\rangle\langle \phi_i|
\end{equation}
where \{$|\phi_i\rangle$\} are the eigenstates of the undriven system. In Fig. \ref{spectrum} are plotted the lowest four energy levels of the system as a function of the reduced gate charge $n_{\mathrm{g}}$. At the particular bias point $n_{\mathrm{g0}}=1/2$, the system is immune to first-order fluctuations of the gate charge and the lowest decoherence rates are obtained \cite{Vion02,Ithier05}.

Manipulation of the quantum state is performed by adding to the dc part of the gate voltage ac microwave pulses with small amplitudes $n_g(t)=n_{g0}+n_g^{ac}(t)$. Adding the ac part of the drive, the Hamiltonian in the basis of the undriven system eigenstates, becomes

\begin{equation}
\label{stirap_quantronium}
H = \sum_i E_{i} |\phi_i\rangle\langle \phi_i|+n_g^{ac}(t)\sum_{i,j} n_{ij}|\phi_i\rangle\langle \phi_j|
\end{equation}
where $n_{ij}=\langle\phi_i|$\^N$|\phi_j\rangle$ are the matrix elements of the number operator in the basis of the undriven eigenstates.

The STIRAP protocol can be carried out between the three lowest energy levels with an ac gate charge $n_g^{ac}=A_g(t)\cos{\omega_g t}+A_u(t)\cos{\omega_u t}$ (see Fig.\ref{spectrum}). Besides the minimal decoherence, at the optimal working point $n_{g0}=1/2$, the microwave field couples off-diagonally in the basis of the eigenstates. However at this point small level spacings and selection rules impede the operation of the scheme \cite{Liu05}. This problem can be circumvented by working slightly away from the optimal point, but not too far to avoid high decoherence. We have chosen $n_{g0}=0.47$ (see Section \ref{sec:2.1} for more details). Here the Hamiltonian is more complicated than the ideal one of Eq. (\ref{stirap_hamilt}), because of the presence of more than three levels, diagonal and counter-rotating drives. Furthermore, the effective peak Rabi frequencies are different (see Sec. \ref{sec:2.3}) and the coupling between the states $|g\rangle$ and $|u\rangle$ is not zero (see Fig. \ref{working_point}a). Nevertheless, if the two ac signals with slightly detuned frequencies $\omega_g$ and $\omega_u$ are applied to the gate, it is possible to adiabatically transfer the population from the ground state $|g\rangle$ to the first excited state $|u\rangle$ just as in a quantum optical system. 

\begin{figure}[h]
\centering
\resizebox{0.4\textwidth}{!}{\includegraphics{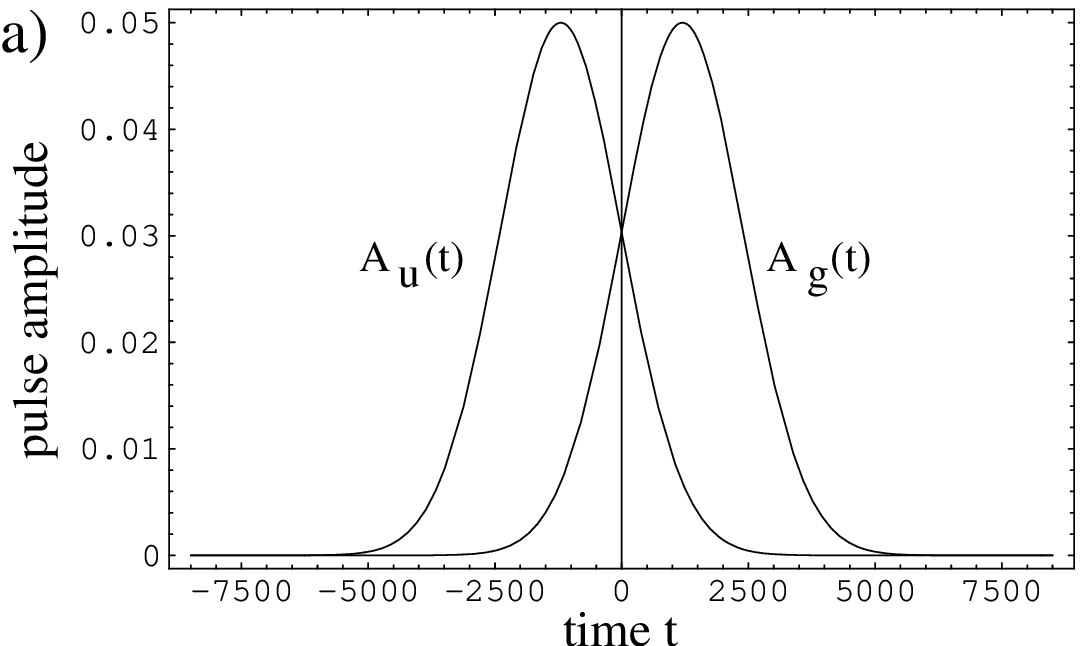}}  \ \ \ \ \ \ \ \ \ \ \
\resizebox{0.4\textwidth}{!}{\includegraphics{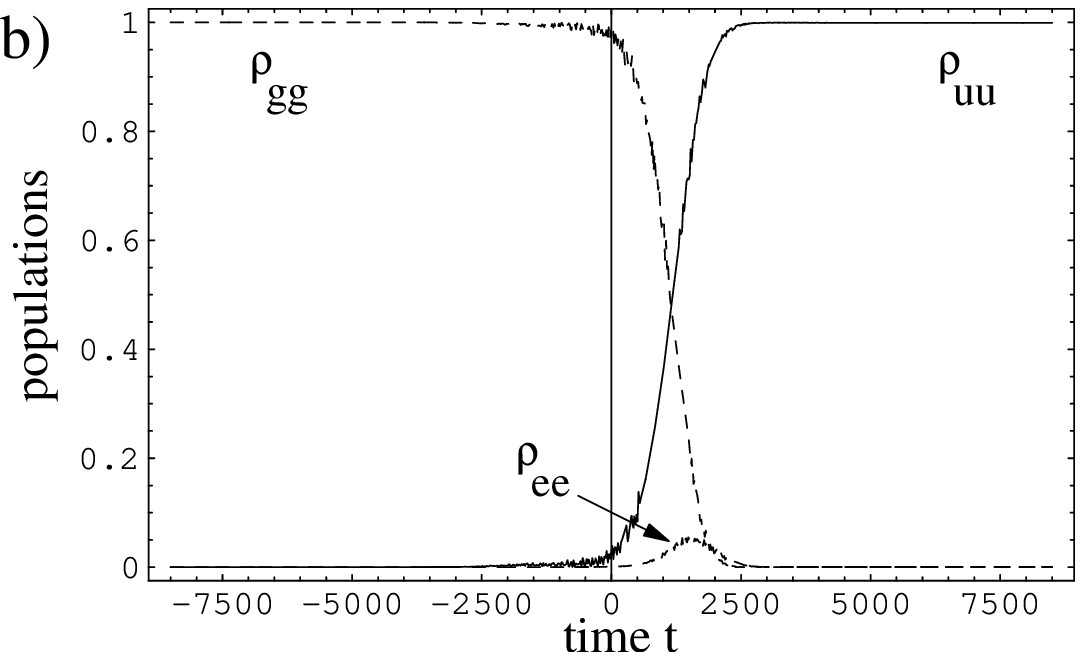}}
 \caption{Population transfer by STIRAP in the quantronium. a) Gaussian-shaped microwave pulses $A_{\mathrm{g}}(t)=A \exp{[-(t-\tau)^2/T^2]}$, $A_{\mathrm{u}}(t)=A \exp{[-(t+\tau)^2/T^2]}$ with a maximum amplitude $A=0.05$. For a charging energy of $E_{\mathrm{C}}=56.9 \mu eV$ the time unit corresponds to about $1.16 \times 10^{-11}$s. The pulse width is $T=20$ns and the delay between the two pulses $\tau=14$ns. Detunings of the frequencies $\omega_{\mathrm{g}}$ and $\omega_{\mathrm{u}}$ are zero.  b) Time evolution of the population $\rho_{\mathrm{gg}}$, $\rho_{\mathrm{uu}}$ and $\rho_{\mathrm{ee}}$ for the isolated system. The efficiency achieved for population transfer is $99.9\%$. }
\label{stirap}
\end{figure}

The von Neumann equation for the isolated system  is
\begin{equation}
 \dot{\rho} = \frac{i}{\hbar}[\rho,H]
\end{equation}
where $\rho$ is the density matrix of the system. In Fig. \ref{stirap}, there we show the numerical solution of this equation for the Hamiltonian (\ref{stirap_quantronium}) truncated to eight charge states. First the system is prepared in the state $|g\rangle$. Then, two Gaussian-shaped microwave pulses are applied in the counterintuitive sequence and finally a population transfer with almost unit efficiency is obtained. The state $|e\rangle$ remains basically unpopulated during the whole procedure as one would expect from the STIRAP protocol.    

\section{Sensitivity to parameters}
\label{sec:2}
Once we have shown how the STIRAP procedure can be implemented in the quantronium, in this section we discuss the parameters which optimize the transfer efficiency.

\subsection{Optimal working point} 
\label{sec:2.1}
\begin{figure}[h]
\centering
\resizebox{0.4\textwidth}{!}{\includegraphics{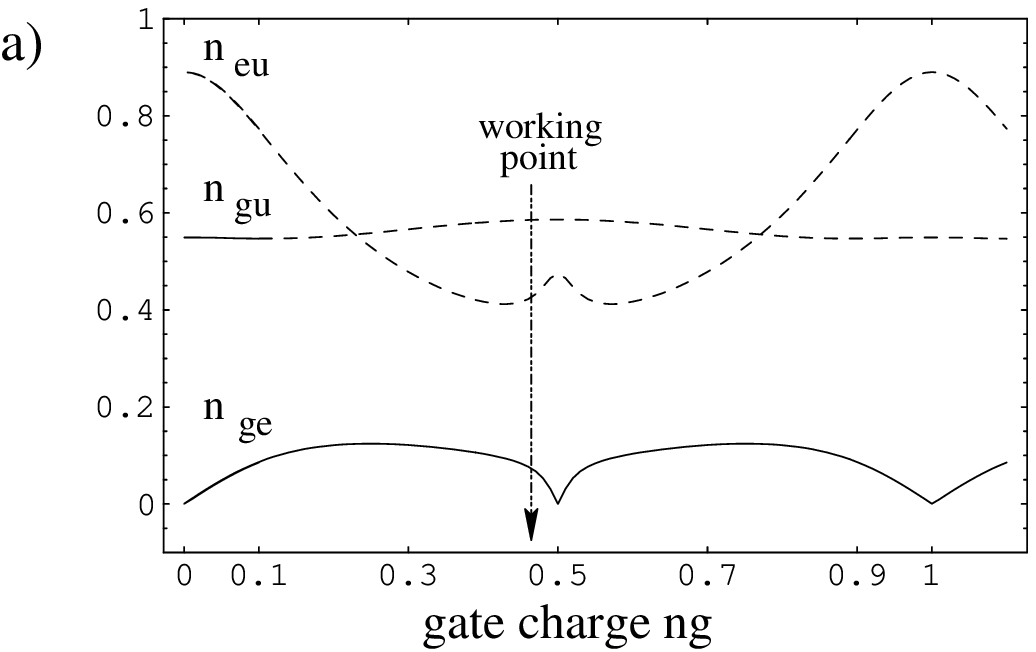}} \ \ \ \ \
\resizebox{0.4\textwidth}{!}{\includegraphics{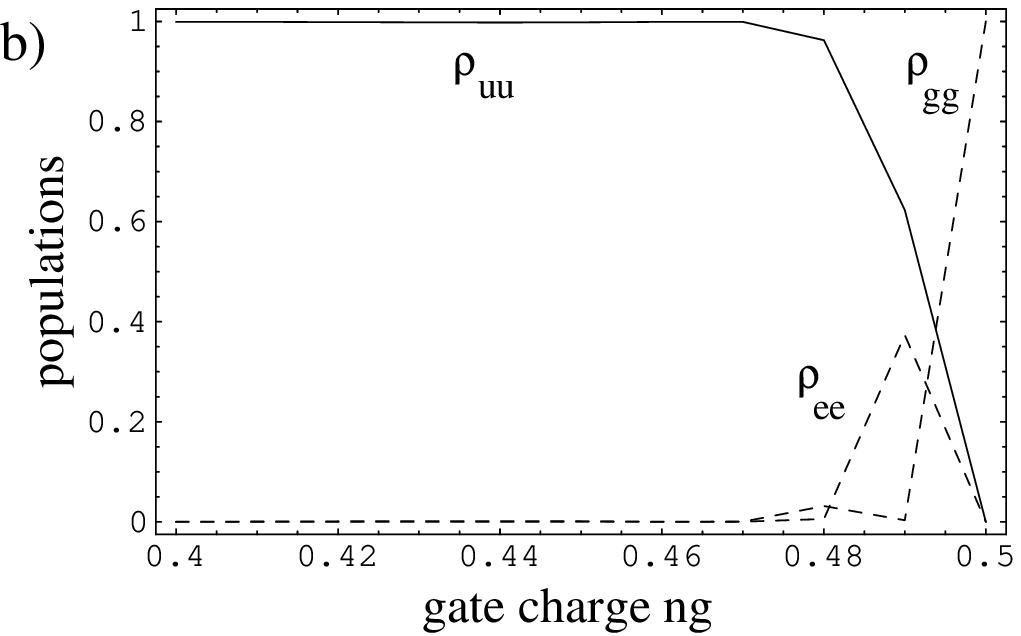}} 
\caption{a) Charge matrix elements for the three levels involved in the STIRAP operation. At the point $n_{\mathrm{g0}}=1/2$, the coupling term $n_{\mathrm{ge}}=0$ due to selection rules. b) Populations $\rho_{\mathrm{uu}}$, $\rho_{\mathrm{ee}}$ and $\rho_{\mathrm{gg}}$ at the end of the procedure as a function of the gate charge. A pulse width $T=20$ns, a pulse delay $\tau=14$ns and Gaussian microwave pulses with maximum amplitude $A=0.05$ have been used. In absence of noise and with the parameters chosen, population transfer is efficient for $n_{\mathrm{g0}}\leq 0.47$.\label{working_point}}
\end{figure}

As described in the previous section, it would be preferable to work at the optimal point $n_{\mathrm{g0}}=1/2$, where decoherence is minimal and the drive is purely off-diagonal. However at this protected point, unwanted transitions to higher levels can occur and, moreover, the parity of the wavefuntions induces selection rules \cite{Liu05}, which render the implementation of STIRAP protocol impossible. In particular the selection rules make the coupling between the states $|g\rangle$ and $|e\rangle$  vanish at the optimal point (see Fig. \ref{working_point}a). The solution, then, is to work away from this point in order to have enough coupling between the above-mentioned states. On the other hand, moving excessively away from the optimal point leads to high decoherence rates and quasi-particle tunneling processes \cite{Ithier05}.

In Fig. \ref{working_point}b, we plot the populations $\rho_{\mathrm{uu}}$, $\rho_{\mathrm{ee}}$ and $\rho_{\mathrm{gg}}$ at the end of the STIRAP procedure as a function of the bias gate charge $n_{g0}$. One can immediately see that at $n_{\mathrm{g0}}=1/2$ the population transfer is not achieved and the system remains in the initial state $|g\rangle$. Moreover, the further we move away from the protected point, the more the transfer efficiency (which coincides with the population $\rho_{\mathrm{uu}}$)  grows due to the increase of the coupling $n_{\mathrm{ge}}$. A good choice which assures high efficiency and acceptable decoherence rates is $n_{\mathrm{g0}}=0.47$.

\subsection{Sensitivity to Delay}
\label{sec:2.2}
The following considerations concern the choice of an optimal delay between the two pulses. The optimal delay, which leads to maximal transfer efficiency, must be chosen by maximizing adiabaticity. The adiabatic condition requires $|\dot{A}_j/A_j|<\omega_j\ (j=u,g)$. This condition translates in an upper bound on the time derivative of the mixing angle $\theta(t)$, which is proportional to non-adiabatic couplings \cite{Bergmann98}. Thus for maximal adiabaticity the mixing angle $\theta(t)$ must change slowly and smoothly in time in order to guarantee small non-adiabatic couplings .

\begin{figure}[h]
\centering
\resizebox{0.5\textwidth}{!}{\includegraphics{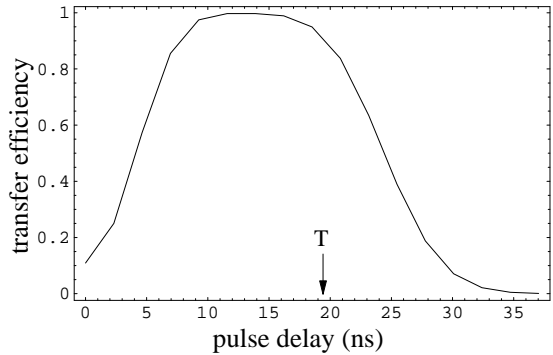}}
 \caption{Transfer efficiency as a function of the pulse delay. The optimal delay is slightly lower than the pulse width $T\approx20$ns for Gaussian-shaped pulses. In the low efficiency regions non-adiabatic couplings between the instantaneous eigenstates occur.\label{opt_delay}}
\end{figure}

The numerical simulation in Fig. \ref{opt_delay} shows that for our choice of Gaussian pulses, the optimum occurs when the delay is slightly lower than the pulse width.

\subsection{Sensitivity to Rabi frequencies}
\label{sec:2.3}
For STIRAP it is better to have two nearly equal peak Rabi frequencies (maximum drive amplitudes). But looking at Fig. \ref{working_point}a, we see that the charge matrix element coupling the states $|g\rangle$ and $|e\rangle$ ($n_{\mathrm{ge}}$) is smaller than the one coupling the states $|e\rangle$ and $|u\rangle$ ($n_{\mathrm{eu}}$). This means that the effective peak Rabi frequencies are different $n_{\mathrm{ge}} A_{\mathrm{g}}<n_{\mathrm{ue}} A_{\mathrm{u}}$. It is interesting to note that, in our analogy with quantum optics, the charge matrix elements correspond to the dipole transition moments.

 If the two maximum Rabi frequencies are different, while the pulse widths are about the same, the projection of the state vector onto the adiabatic transfer state is very good initially (because in our case the more intense drive comes first), but necessarily less good in the final stage and consequently the transfer efficiency will be small.

It is also true that using large pulse areas, small deviations from the optimal conditions do not lead to significant drop in transfer efficiency. In fact adiabaticity requires a large pulse area \cite{Bergmann98} and non-adiabatic transitions can be suppressed by increasing the drive amplitudes. However using equal peak Rabi frequencies and equal pulse widths allows to reduce the necessary pulse areas and facilitate efficient population transfer. Furthermore in solid state qubits it is not easy to increase the microwave pulse amplitude beyond a certain value since other excitations come into play. Thus it is important to externally adjust the drive amplitudes in such a way that the effective Rabi frequencies are about the same ($n_{\mathrm{ge}} A_{\mathrm{g}}\approx n_{\mathrm{ue}} A_{\mathrm{u}}$). 

\begin{figure}[h]
\centering
\resizebox{0.5\textwidth}{!}{\includegraphics{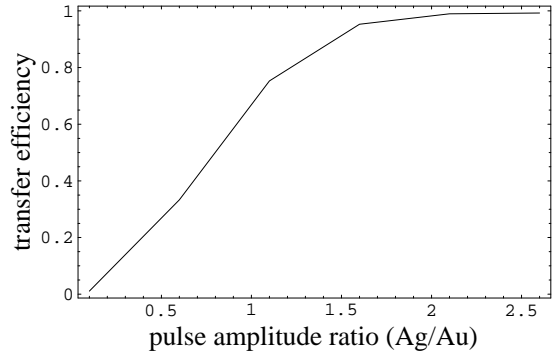}}
 \caption{Transfer efficiency as a function of the ratio of the maximum drive amplitudes. The simulation has been done with a pulse width $T=20$ns, a pulse delay $\tau = 14$ns. The ratio of the Rabi frequencies has been changed by keeping fixed the amplitude $A_{\mathrm{u}}=0.02$ and varying $A_{\mathrm{g}}$. The value $A_{\mathrm{g}}/A_{\mathrm{u}}= 2$ assures a high transfer efficiency .\label{rabi_frequencies}}
\end{figure}

In Fig. \ref{rabi_frequencies}, we report a numerical plot of the transfer efficiency as a function of the ratio of the two peak Rabi frequencies. Simply making the drive $A_{\mathrm{g}}$ twice as strong as the drive $A_{\mathrm{u}}$, we get a high transfer efficiency with reasonable amplitudes.

\subsection{Sensitivity to detunings}
\label{sec:2.4}
When the two frequencies $\omega_{\mathrm{g}}$ and $\omega_{\mathrm{u}}$ are not exactly resonant with the respective transitions, the presence of non-zero detunings can affect the efficiency. There are two cases of interest. First, both $\omega_{\mathrm{g}}$ and  $\omega_{\mathrm{u}}$ may vary, while the two-photon resonant condition is maintained (single-photon detuning, $\delta=0$). Alternatively either frequency may vary, while the other is fixed (two-photon detuning, $\delta\ne0$).

The single-photon detuning does not affect the formation of the dark state, because the mixing angle does not depend on it. However this type of detuning affects the adiabatic condition \cite{Bergmann98} and when we increase $\delta_{\mathrm{g}}$ (and consequently $\delta_{\mathrm{u}}$), the transfer efficiency decreases for deterioration of adiabaticity. 

The detuning from two-photon resonance is more detrimental for STIRAP, because it prevents the exclusive population of the trapped state, which is no longer an instantaneous eigenstate of the Hamiltonian. A more detailed analysis of the instantaneous eigenstates when $\delta\ne0$ shows that there is no adiabatic transfer state providing an adiabatic connection from the initial to the target state, as does the dark state for $\delta=0$ and the evolution leads to complete population return of the system to its initial state. The only mechanism which leads to population transfer  is by non-adiabatic transitions between the adiabatic states \cite{Bergmann98}. Actually for small values of $\delta$, narrow avoided crossings between the instantaneous eigenvalues can occur and the population can be transferred by Landau-Zener tunneling \cite{Landau32}.

\begin{figure}[h]
\centering
\resizebox{0.0335\textwidth}{!}{\includegraphics{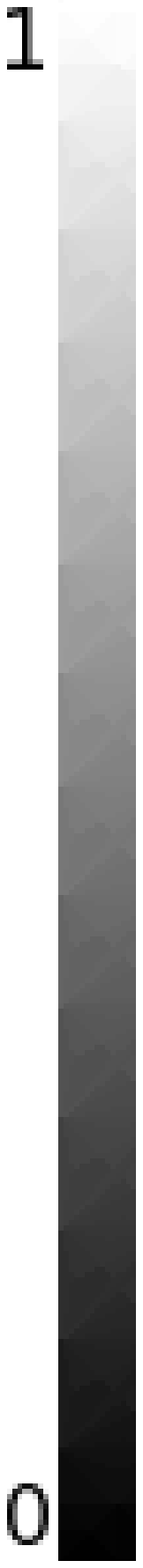}}\ \ \ 
\resizebox{0.4\textwidth}{!}{\includegraphics{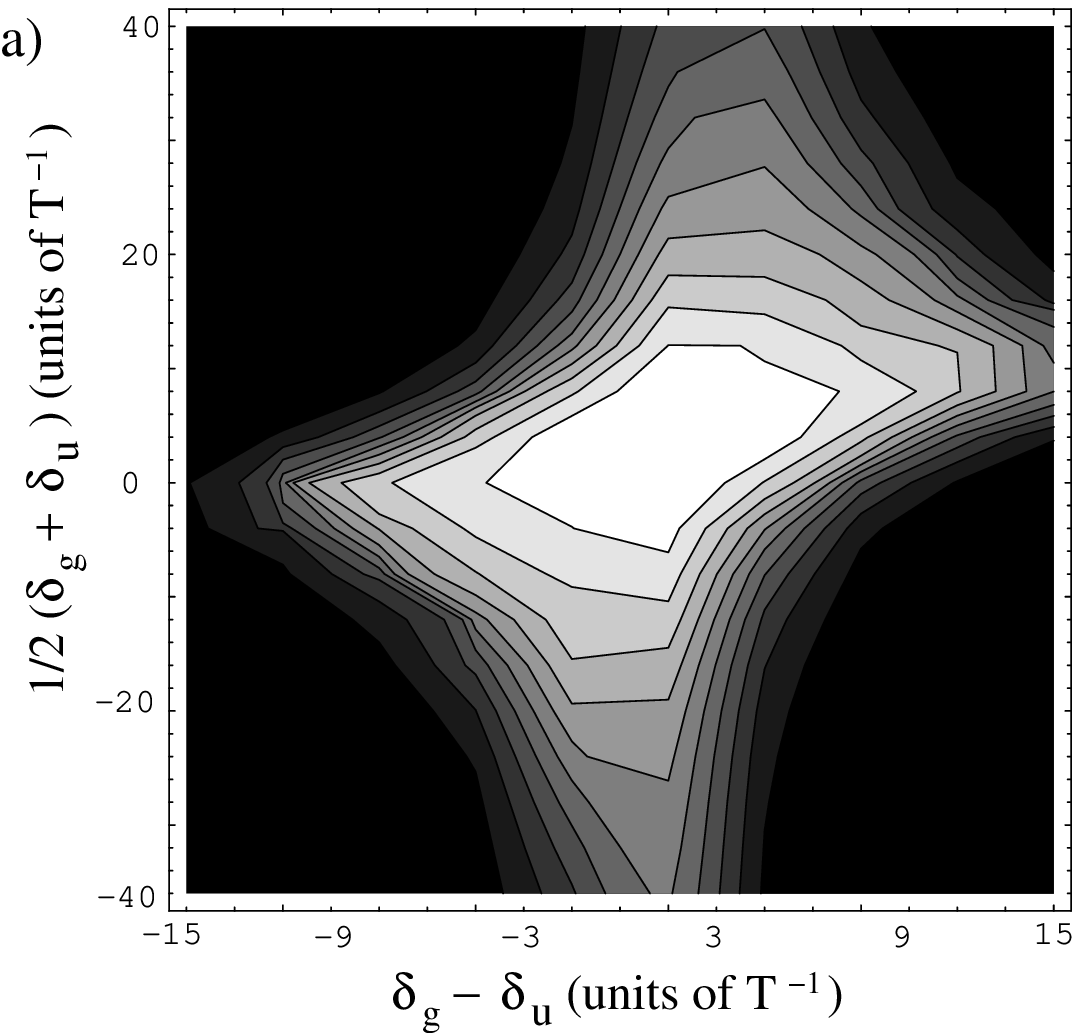}}\ \ \ \
\resizebox{0.4\textwidth}{!}{\includegraphics{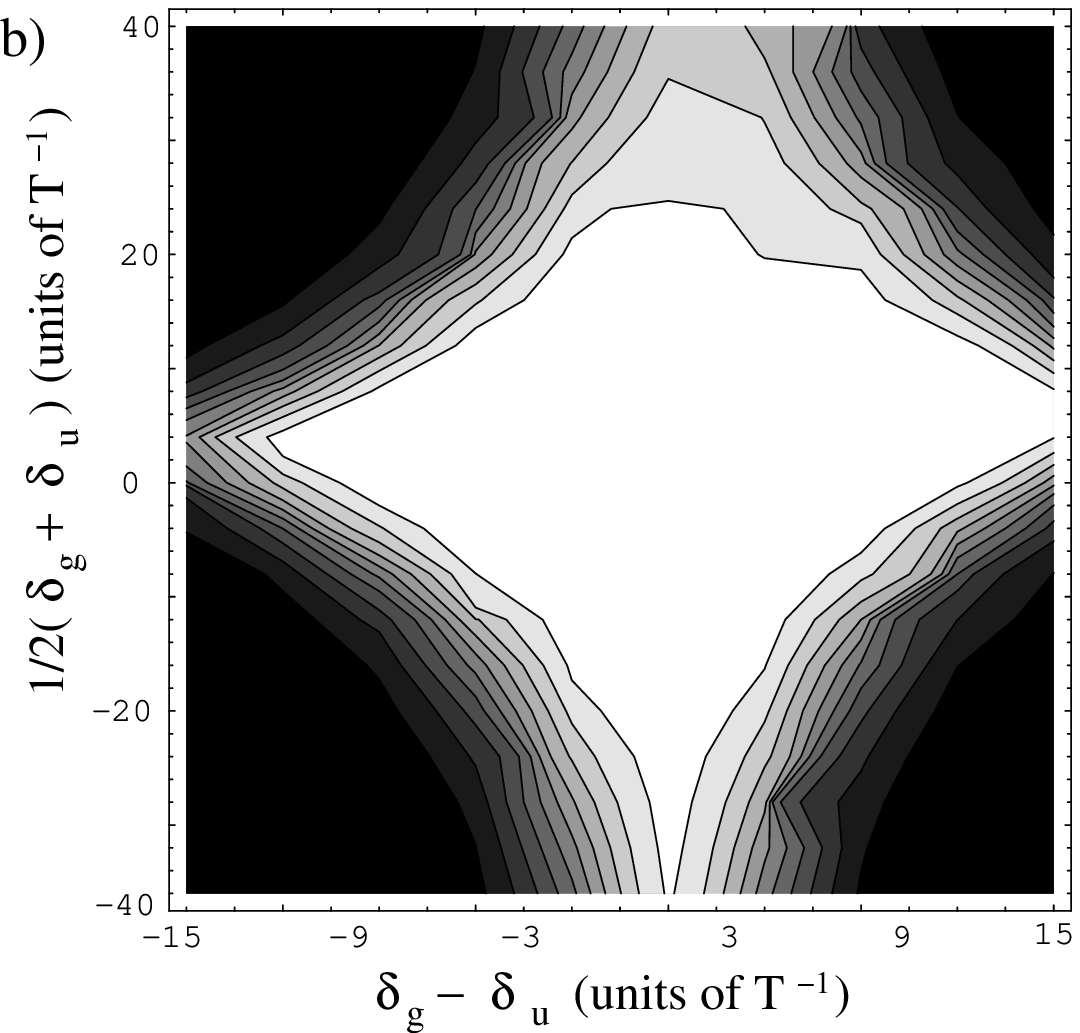}}
\caption{Transfer efficiency as a function of single-photon and two-photon detuning for equal peak
 Rabi frequencies a) ($A_{\mathrm{g}}=A_{\mathrm{u}}=0.03$)  and for externally adjusted amplitudes b) ($A_{\mathrm{g}}=2 A_{\mathrm{u}}=0.06$). The region of high efficiency is larger in the figure b) showing the importance of choosing appropriately adjusted external amplitudes.\label{detunings}}
\end{figure}

To show how STIRAP efficiency is much less sensitive to single photon detuning than to two-photon detuning, we show in Fig. \ref{detunings} the numerically calculated transfer efficiency plotted versus the single-photon and the two-photon detuning. The figure shows two plots of the efficiency sensitivity to detunings for equal drive amplitudes (Fig. \ref{detunings}a) and for externally adjusted drive amplitudes (Fig \ref{detunings}b). In the case of adjusted peak Rabi frequencies (see section \ref{sec:2.3}), the region of high efficiency is larger. This is important for reducing the effect of low frequency noise treated in section \ref{sec:3.2}.

\section{Effects of decoherence}
\label{sec:3}
The efficiency of solid-state quantum-coherent nanodevices is affected by various (device dependent) noise sources. In the quantronium, high-frequency noise and low-frequency noise coexist. The former is mainly responsible for unwanted transitions which lead to exponential decay of the signal in time, whereas the latter mainly makes the calibration of the device unstable and determines the initial power-law decay of the signal \cite{Paladino02,Ithier05}.

Based on the above considerations, we discuss the feasibility of the protocol without making a detailed analysis of decoherence in the protocol \cite{Stirap_decoherence}, which is beyond the scope of this work. The key observation is that strong unwanted processes involving the state $|e\rangle$ have negligible effects both for high and for low-frequency noise on the STIRAP protocol. Decoherence is mainly determined by processes involving the states $|g\rangle$ and $|u\rangle$, which have been well characterized in the quantronium and, as a matter of fact, allow decoherence times $\tau_{\mathrm{R}} > 300 \,ns$\cite{Ithier05}.

\subsection{High frequency noise}
\label{sec:3.1}
\begin{figure}[h]
\centering
\resizebox{0.5\textwidth}{!}{\includegraphics{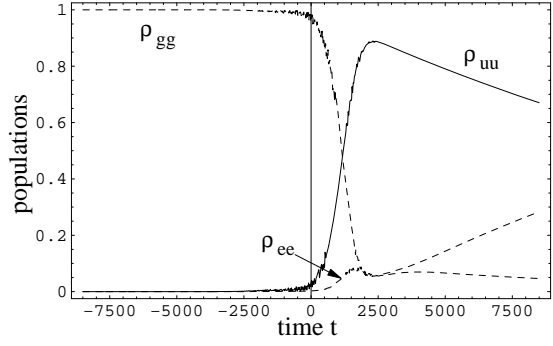}}
 \caption{Time evolution of the populations $\rho_{gg}$, $\rho_{ee}$ and $\rho_{uu}$ with decoherence. The time unit corresponds to about $1.16 \times 10^{-11}$s ($ h / E_{\mathrm{C}}$). For the calculations with quantum noise we have used the decay rate $\gamma_{\mathrm{u}}=4.4\times10^{-5}$ and the dephasing rate $\tilde{\gamma}=2.6\times10^{-4}$ \cite{Ithier05}. The other parameters are those used in Fig. \ref{stirap}. The extra decay $\gamma_{\mathrm{u}\to\mathrm{g}}$ acts only at the end of the procedure and the increase in the population of level $|e\rangle$ is harmless.
\label{high_freq_noise}}
\end{figure}

High frequency noise can be analyzed by the quantum-optical master equation
\begin{equation}
\label{master_eq}
 \dot{\rho} = \frac{i}{\hbar}[\rho,H^{\prime}] - \Gamma\rho
\end{equation}
where $\rho$ is the density matrix of the system and $H^{\prime}$ is the Hamiltonian (\ref{stirap_quantronium}) in the rotating frame \cite{Kuhn99}. The dissipator $\Gamma\rho$ includes spontaneous decay rates and environment-assisted absorption in the presence of the driving fields. In the basis $\{|g\rangle,|e\rangle,|u\rangle\}$, it reads
\begin{equation}
\label{dissipator}
(\Gamma\rho)_{ij} = \frac{\gamma_i+\gamma_j}{2}
                                          \rho_{ij}
         -(1-\delta_{ij})\tilde{\gamma}\rho_{ij}
         -\delta_{ij}\sum_k \rho_{kk}
                                     \gamma_{k\to i}
\end{equation}
where $\gamma_i=\sum_{k\neq i} \gamma_{\mathrm{i}\to \mathrm{k}}$. We take the dissipator time-independent (which overestimates decoherence) and include all transitions as well as dephasing rate $\tilde{\gamma}$, which accounts phenomenologically for low-frequency noise. For the state $|e\rangle$ (the second excited state of the quantronium) we assume $\gamma_{\mathrm{e}}=\gamma_{\mathrm{e}\to\mathrm{u}}+                \gamma_{\mathrm{e}\to\mathrm{g}}=2\gamma_{\mathrm{u}}$ where $\gamma_{\mathrm{u}}$ is taken on the order of the rate observed in the experiments of Ref. \cite{Ithier05}. While in quantum optics the rate $\gamma_{\mathrm{u}\to\mathrm{g}}$ vanishes and the remaining decay rates act on depopulated states hardly affecting the protocol, STIRAP in the quantronium may be sensitive to the extra decay $|u\rangle\to|g\rangle$ involving the two low-lying states. 
The Fig. \ref{high_freq_noise} show results for the master equation (\ref{master_eq}). It can be recognized a good robustness of the procedure against decoherence. We observe that the extra rate $\gamma_{\mathrm{u}\to\mathrm{g}}$ changes the population only during the waiting time after completion of the pulse sequence and the population of level $|e\rangle$ is slightly increased.

\subsection{Low frequency noise: detuning instability}
\label{sec:3.2}
\begin{figure}[h]
\centering
\resizebox{0.6\textwidth}{!}{\includegraphics{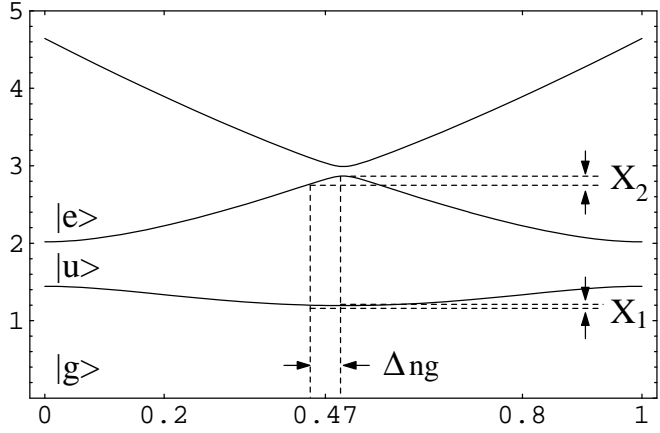}}
 \caption{Detuning instability due to low-frequency noise. Fluctuations in the gate charge $\Delta n_{\mathrm{g}}$ translate in fluctuating single-photon detuning $\delta_{\mathrm{g}}\rightarrow \delta_{\mathrm{g}}+X_2$ and fluctuating two-photon detuning $\delta \rightarrow \delta+X_1$.
\label{low_freq_noise}}
\end{figure}

Low frequency noise is modelled as due to impurities which can be considered static during each run of the protocol but switch on a longer time scale, thus leading to a statistic distribution of level separations. In protocols like Ramsey interference, they result in a distribution of oscillating frequencies, whose average determines defocusing of the signal \cite{Paladino02}. The effect of averaging over static impurities configurations corresponds to statistically distributed energy level fluctuations $\delta E_{\mathrm{i}}$, which, in the rotating frame, translates in fluctuations of the detunings. As one can see from Fig. \ref{low_freq_noise}, even large fluctuations of $E_{\mathrm{e}}$ ($X_2$) hardly affect STIRAP, since they still leave equal detunings of both microwave fields (i.e. they do not affect two-photon resonance, see section \ref{sec:2.4}). On the other hand, fluctuations of the difference between the energies of the two lowest eigenstates $E_{\mathrm{u}}-E_{\mathrm{g}}$ (X1) are potentially detrimental since they translate in two-photon detuning fluctuations (see section \ref{sec:2.4}). However as long as the system is well-protected from noise, fluctuations remain inside the high efficiency region in Fig. \ref{detunings}  and  almost complete population transfer can be achieved. 

\section{Conclusions}
\label{conclusion}
In summary  we have discussed the possibility to implement STIRAP in the quantronium device. One important advantage of the protocol is that the efficiency does not depend sensitively on details of the procedure and on timing making it robust against moderate fluctuations of the solid-state environment. One goal of this work is to provide a simple basis and parameters for the experimental realization of STIRAP. Since we have used real circuit design with common parameters and corresponding time scale, it should be feasible to experimentally verify STIRAP in nanocircuits with the state-of-the-art technology. 

There are many interesting applications of this scheme such as the preparation of Fock states and the single photon generation\cite{Kuhn99} in cavity coupled to the nanocircuit. The cavity can be implemented using electrical resonator \cite{Falci03}, transmission lines \cite{Wallraff04} and nanomechanical resonators \cite{Siewert05}.

\end{document}